\documentclass[11pt]{article}
\usepackage{Blois,epsfig}

\def\Journal#1#2#3#4{{#1} {\bf #2}, #3 (#4)}

\def\NPB{{\em Nucl. Phys.} B}
\def\PLB{{\em Phys. Lett.}  B}

\def\be{\begin{equation}}
\def\ee{\end{equation}}
\def\bea{\begin{eqnarray}}
\def\eea{\end{eqnarray}}

\newcommand{\mbf}[1]{\mbox{\boldmath $#1$}}
\newcommand{\brho}{\mbf{\rho}}
\begin{document}

\vspace*{2cm}
\title{Representations of the LL BFKL Kernel and the Bootstrap}
\author{G.P. Vacca}
\address{INFN and Dipartimento di Fisica, Via Irnerio 46,
I-40126 Bologna, Italy\\
e-mail: vacca@bo.infn.it}
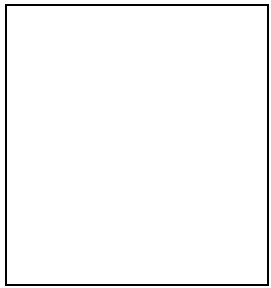
\maketitle\abstracts{
Different forms of the BFKL kernel both in coordinate and momentum
representations may appear as a result of different gauge choices and/or inner
scalar products of the color singlet states. We study a spectral representation
of the BFKL kernel not defined on the M\"obius space of functions but on a
deformation of it, which provides the usual bootstrap property due to
gluon reggeization. In this space the corresponding symmetry is
made explicit introducing a deformed realization of the $sl(2,C)$ algebra.}
\section*{Introduction}
Small $x$ physics in perturbative QCD has been long investigated, after the
pioneering work in the LL approximation devoted to the study of the
asymptotical behavior in the high energy limit of the physical cross
sections~\cite{BFKL}.
The kinematics of such processes is related to the Regge limit wherein one
observes the high energy factorization of the scattering amplitudes in terms of
external particle impact factors and a Green's function, which exponentiates
the BFKL kernel, describing the evolution in rapidity of the states in the
t-channel.

The leading energy contribution to a perturbative cross section is
typically associated to the exchange of the BFKL Pomeron,
which in the lowest approximation is realized as the exchange of a pair of
two interecting reggeized gluons in the color singlet state.
The color singlet impact factors by gauge invariance vanish as one of the two
reggeized gluons carries a zero transverse momentum so that the
BFKL kernel (Green's function) is not unique: it depends for example on the
gauge choices, and therefore one may choose different space of functions
for the particle impact factors and the eigenstates which diagonalize the
BFKL kernel dynamics.

Not less important is the color octet exchange channel: the kernel
leads to an equation which has a bootstrap solution, a fundamental consistency
property derived from the s-channel unitarity and corresponding to the
reggeization of the gluon. This property can be
formulated also in the NLL approximation~\footnote{see Fadin's
contribution at EDS05}.
Different bootstraps conditions can be defined
but the most important is again the one directly related to the gluon
reggeization (in the strong form), telling that all the colored impact
factors are proportional to the gluon eigenstate of the octet
kernel~\cite{StrongB}. 

In the LL approximation the impact factors of the colored external particles
are simply functions only depending on the total transverse momenta exchanged in
the t-channel. 
A manifestation of this
bootstrap property also takes place for the color singlet state
of two gluons and inside the coupling of three or more gluons to
colorless particle impact factors. In particular, it plays an
important role in the Odderon solution~\cite{BLV} which appears
as a bound state of three reggeized gluons and in generalizations of
it~\cite{vacca}. 

After the construction of the BFKL kernel from a Feynman diagram analysis one
faces two cases.
For an amplitude related to the exchange of two reggeized gluons in the singlet
state, due to the gauge invariance, one may move to the so called M\"obius space
of functions~\cite{Lipatov86,BLV2}; then the kernel (which acts on amputated
functions) presents remarkable
properties such as holomorphic factorization and M\"obius symmetry.
The latter is exploited in coordinate representation and
the BFKL kernel can be easily diagonalized on the conformal eigenstate basis
associated to the famous Polyakov ansatz.

On the other hand one is not allowed to do this when the two interacting
reggeized gluons are in an octet state. The relevant kernel simply acts
on a different space of functions.
In the latter the singlet kernel inherits a property from
bootstrap but the M\"obius symmetry is no more so evident.
It has been shown~\cite{BLSV} that the M\"obius symmetry $sl(2,C)$
is still a property of the singlet kernel in such a space (of non amputated
functions) and that its realization is deformed.
According to this one may still write a spectral , but deformed,
representation of the BFKL kernel, which can be also useful in the study of
coupling the BFKL kernel directly to quarks~\cite{MT,BFLLRW}.
\section*{LL BFKL kernels, bootstrap and deformation}
In the following we shall illustrate in few steps the main result leaving
out much of the details which can be found elsewhere~\cite{BLSV}.  

Let us start from the BFKL kernel~\cite{BFKL,Lcft} operator which generates
the Green's function appearing in the high energy factorization of an amplitude:
\begin{equation}
K_2^{(R)}= -
\left( \omega_1+ \omega_2 \right) - \lambda_R V_{12}\,,
\label{bfklkernel}
\end{equation}
where R labels the colour representation of the two gluon state and in the
singlet and octet channel one has respectively $\lambda_1=N_c$ and $%
\lambda_8=N_c/2$. The $\omega$'s are trajectories of the reggeized
gluons (virtual contributions) and $\lambda_R V_{12}$ are the real
corrections due to the gluon emission strongly ordered in rapidity (Multi
Regge Kinematics).
The BFKL operator $K_2$ is initially defined in momentum space and we do not
need the explicit expressions here.
In the octet channel the bootstrap condition reads:
\begin{equation}
\bar{K}_{12}^{(8)} \otimes 1=-\omega(\mbox{\boldmath $q$})\, ,
\label{bootstrap}
\end{equation}
where the barred operator is chosen to be amputated on the left and $\otimes$
makes explicit the integration in momentum space.
This relation implies on the singlet case the relation
\begin{equation}
\bar{K}_{12}^{(1)}\otimes 1=-2 \omega(\mbox{\boldmath $q$}) +\omega(%
\mbox{\boldmath $k$}_1)+\omega(\mbox{\boldmath $k$}_2)=
 \frac{1}{2} \bar{\alpha}_s \log\left( \frac{ \mbox{\boldmath
$q$}^4}{\mbox{\boldmath $k$}_1^2 \mbox{\boldmath $k$}_2^2}
\right)\,,
\label{bootstrap2}
\end{equation}
with $\bar{\alpha}_s=\alpha_s N_c/\pi$, which is infrared finite in $4$ dimensions.

In the singlet case because of gauge invariance an impact factor in momentum
space is such that
$\Phi(\mbf{k}_1,\mbf{k}_2) \to 0$ as $\mbf{k}_i \to 0$ and therefore we can
add arbitrary terms, proportional to $\delta^{(2)}(\mbf{k}_i)$,
to the gluon propagators and to the BFKL kernel,
since they do not alter the amplitude.
In coordinate space this corresponds to the physical equivalence relation
\begin{equation}
f(\mbox{\boldmath $\rho $}_{1},\mbox{\boldmath $\rho $}_{2}) \sim \tilde{f}(
\mbox{\boldmath $\rho $}_{1},\mbox{\boldmath $\rho $}_{2}) = f(
\mbox{\boldmath $\rho $}_{1},\mbox{\boldmath $\rho $}_{2})+ f^{(1)}(
\mbox{\boldmath $\rho $}_{1})+f^{(2)}(\mbox{\boldmath $\rho $}_{2})\,,
\label{uv_transf}
\end{equation}
which permits to choose to work in the space of M\"obius functions
$f(\mbox{\boldmath $\rho $}_{1},\mbox{\boldmath $\rho $}_{2})$, such that
$f(\mbox{\boldmath $\rho $},\mbox{\boldmath $\rho $})=0$.

In such a space the operator $H_{12}=K_{12}^{(1)}\,2 /\bar{\alpha}_s $ is
M\"obius invariant and can be written in the separable form
\begin{equation}
H_{12}=h_{12}+h_{12}^{\ast },\,\,h_{12}=\sum_{r=1}^{2}\left( \ln p_{r}+\frac{%
1}{p_{r}}\ln (\rho _{12})\,p_{r}-\Psi (1)\right) \,,  \label{BFKL_sep}
\end{equation}
where one may verify the invariance under action of the generators:
\begin{equation}
M_{r}^{3}=\rho _{r}\partial _{r}\,,\,\,M_{r}^{+}=\partial
_{r}\,,\,\,M_{r}^{-}=-\rho^2_{r}\partial _{r}\, .  \label{moebius_gen}
\end{equation}
For two reggeized gluons one has $M^k=\sum_{r=1}^2 M^k_r$ and the Casimir
operator is defined by
$M^2 = |\vec{M}|^2=-\rho _{12}^2\,\partial _1\,\partial _2$ and shares
the eigenstates with the BFKL kernel of eq. (\ref{BFKL_sep}):
\begin{equation}
E_{h,\bar{h}}(\brho_{10},\, \brho_{20}) \equiv \langle \rho | h \rangle =\left(%
\frac{ \rho _{12}}{\rho _{10}\rho _{20}}\right) ^{h}\left( \frac{\rho
_{12}^{*}}{ \rho _{10}^{*}\rho _{20}^{*}}\right) ^{\bar{h}}\,.
\label{pomstatescoord}
\end{equation}
The BFKL kernel on the M\"obius space can also be written in the {\em dipole}
picture~\cite{dipole,BLV2} form
\begin{equation}
H_{12}\,f_{\omega }(\mbox{\boldmath $\rho$}_{1},\mbox{\boldmath $\rho $}%
_{2})= \int \frac{d^{2}\brho _{3}}{\pi }\,\frac{\left| \brho _{12}\right| ^{2}%
}{\left| \brho _{13}\right| ^{2}\left| \brho _{23}\right| ^{2}}\,\left(
f_{\omega }(\mbox{\boldmath $\rho $}_{1},\mbox{\boldmath $\rho $}%
_{2})-f_{\omega }(\mbox{\boldmath $\rho $}_{1},\mbox{\boldmath $\rho $}%
_{3})-f_{\omega }(\mbox{\boldmath $\rho $}_{2},\mbox{\boldmath $\rho $}%
_{3})\right) \,.
\end{equation}
The $sl(2,C)$ (M\"obius) symmetry allows us to write a spectral representation
for the kernel
\begin{equation}
\langle \rho |\hat{\bar{K}}_{12}^{(1)}| \rho^{\prime}\rangle
= \int d^2 \!\mbox{\boldmath $\rho$}_0 \sum_h \frac{N_h}{%
|\brho_{12}|^4} \, \langle \rho | h \rangle \, \chi_h \, \langle h | \rho'
 \rangle\,. 
\end{equation}
This representation is not compatible with the eq. (\ref{bootstrap2}), since
the state $|h\rangle$ is ortogonal to any octet impact factor~\cite{BLSV}
(depending in momentum rpresentation only on the total momentum $\mbf{q}$). 

On noting that the Fourier tranform of the Pomeron eigenstates,
$\langle \mbf{k} | h \rangle=\langle \mbf{k} | h^A \rangle+
\langle \mbf{k} | h^\delta \rangle$ is decomposed in the sum of two
terms~\cite{BBCV}, the first meromorphic in the momenta and the second with
$\delta(\mbf{k}_i)$ singularities, one can trace the appearance of the latter
because of the gauge choice, made to work on the M\"obius space of functions.  
Infact the original kernels in eq. (\ref{bfklkernel}) are meromorphic in the
momenta. 

It is therefore natural to define a completeness relation~\cite{BLSV} such that
\begin{equation}
\langle \mbf{k} |\hat{\bar{K}}_{12}^{(1)}| \mbf{k}^{\prime}\rangle  = \sum_h \tilde{N}_h
\,\langle \mbf{k} | h^A \rangle 
\, \chi_h \, \langle h^A | \mbf{k}' \rangle \,,  \nonumber
\end{equation}
which is associated to the scalar product
(corresponding to the Analytic Feynman (AF) space)
\begin{equation}
\langle f | g \rangle \equiv \int d\mu(\mbox{\boldmath $k$}) f^*(%
\mbox{\boldmath $k$}_1,\mbox{\boldmath $k$}_2) g(\mbox{\boldmath $k$}_1,%
\mbox{\boldmath $k$}_2) \,, \quad 
d \mu(\mbox{\boldmath $k$}) =\mbox{\boldmath $k$}_1^2 \mbox{\boldmath $k$}%
_2^2 \, \,\delta^2(\mbox{\boldmath $q$}-\mbox{\boldmath $k$}_1-%
\mbox{\boldmath $k$}_2) \, d^2\mbox{\boldmath $k$}_1 d^2\mbox{\boldmath $k$}%
_2  \,.
\label{scal_prod}
\end{equation}
This representation of the kernel gives the same result of the M\"obius one
when acting on colorless impact factors but it can also act on colored impact
factors and in particular it satisfies the relation in eq. (\ref{bootstrap2}).
In order to prove this fact we project the above relation on the spectral
basis,
\begin{equation}
\epsilon_h \, \left(\langle h^{A} | P_\lambda \rangle\right)_{\lambda=0} =
\frac{1}{2}\langle h^{A} |   \frac{d}{d
\lambda} \left( | P_\lambda \rangle \right)_{\lambda=0}
\rangle \, , \quad 
\langle k |P_\lambda \rangle \equiv \frac{1}{\mbox{\boldmath $k$}_1^2
\mbox{\boldmath $k$}_2^2} \left( \frac{\mbox{\boldmath $q$}^4}{
\mbox{\boldmath $k$}_1^2 \mbox{\boldmath $k$}_2^2} \right)^ \lambda \, ,
\label{coeff_bootstrap}
\end{equation}
where $\epsilon_h$ is the scaled BFKL kernel eigenvalue.
This new relation can be verified by direct integration~\cite{BLSV} in
momentum space.

It is interesting to understand better the relation between the two spaces
under consideration (with the particular choice made above for the scalar
product in the AF space) and which are spanned by the two basis
$|h\rangle$ (M) and $|h^A \rangle$ (AF). 
Let us write the mapping between the two spaces in coordinate representation:
\begin{eqnarray}
\Phi^{-1} :\mathrm{M \to AF}\!\!\!\! &,&\!\!\!
E^A_h(\brho_{10},\brho_{20})=E^M_h(\brho_{10},\brho_{20})-
\lim_{\rho_1\to\infty} E^M_h(\brho_{10},\brho_{20})-\lim_{\rho_2\to\infty}
E^M_h(\brho_{10},\brho_{20})  \nonumber \\
&{}&\!\!\!\!\!\!=E^M_h(\brho_{10},\brho_{20})-2^{-h-\bar{h}}
\left(E^M_h(\brho_{10},-\brho_{10})+ E^M_h(-\brho_{20},\brho_{20})\right)\,,
\label{mapping_on_E} \\
\Phi :\mathrm{AF \to M :}\!\!\!\! &,&\!\!\!
E^M_h(\brho_{10},\brho_{20})=E^A_h(\brho_{10},\brho_{20})+\frac{1}{2^{h+\bar{h}%
}-2} \left( E^A_h(\brho_{10},-\brho_{10})+
E^A_h(-\brho_{20},\brho_{20})\right)\,.  \nonumber
\end{eqnarray}
This mapping is therefore related to a special shift, defined by its
projection on any basis vector.
One can easily check that
\begin{equation}
\Phi \Phi^{-1} \equiv I_M \, , \quad \Phi^{-1} \Phi \equiv I_{AF}
\label{inverse}
\end{equation}
so that one can really define, by means of this 1-1 mapping, the action of an
operator given on one space also on the other space.
In particular, by this similarity transformation we can define the action of
the M\"obius group generators on the AF space:
\begin{equation}
\vec{M}^{AF}_r=\Phi^{-1} \vec{M}_r \Phi \,.  \label{deREP}
\end{equation}
An explicit realization of these operators can be found~\cite{BLSV} and it
turns out to be simpler to derive it from consideration in the momentum
representation.
Only the $M^-$ generator  has a form different in the M and in the AF spaces
and consequently also the Casimir is different.
Let us note that with other choicea of the scalar product also the
consideration changes. For example considering functions amputated implies that
no $\delta$-like behavior is present and one has meromorphic functions
from the beginning.

One interesting general fact which can be inferred is that the freedom in the
gauge and scalar product choices can make a symmetry more or less evident
and realized in different ways on the associated different space of functions. 

\end{document}